\documentclass[preprint,aps]{revtex4}
\usepackage{xcolor}
\usepackage{graphicx}
\newcommand{\be}{\begin{eqnarray}}
\newcommand{\ee}{\end{eqnarray}}
\usepackage{hyperref}
\usepackage{cancel} 
\newcommand{\nn}{\nonumber}

\newcommand{\ms}{\mskip 1.5mu}

\newcommand{\half}{{\textstyle\frac{1}{2}}}

\usepackage{subfigure}

\newcommand{\clr}[1]{^{#1\hspace{-1px}}}

\usepackage{amsmath,amssymb}

\newcommand{\vek}[1] {\boldsymbol{#1}}

\newcommand{\y}{\vek{y}}
\newcommand{\ytilde}{\tilde{\vek{y}}}

\begin{document}
\preprint{NIKHEF 2016-027}

\title{Quark-gluon double parton distributions in the light-front dressed quark
model}

\author{\bf Tomas Kasemets$^a$ and  Asmita Mukherjee$^b$}

\affiliation{$^a$ Nikhef and Department of Physics and Astronomy, VU
  University Amsterdam, De Boelelaan 1081, 1081 HV Amsterdam, The
  Netherlands\\ $^b$ Department of Physics,\\
Indian Institute of Technology Bombay,\\ Powai, Mumbai 400076,
India.}
%\date{\today}

\begin{abstract}
\noindent
We study parton distributions for two partons, a quark and a gluon, in the light-front dressed quark model, with focus on correlations between the two partons. The model calculation leads to sizable spin-spin and spin-kinematic correlations of interest for studies of double parton scattering (DPS) in high-energy collisions. In particular, we find that the transverse dependence of the double parton distributions (DPDs) does not factorize within the model. The results gives insight to the strengths of correlations in different kinematical regions, which can help in constructing input DPDs in cross section calculations.
\end{abstract}

\maketitle

%%%%%%%%%%%%%%%%%%%%%%%%%%%%%%%%%%%%%%%%%%%%%%%%%%%%%%%%%%%%%%%%%%%%
\section{Introduction}
%%%%%%%%%%%%%%%%%
Multiparton interactions, where more than one parton from each hadron collide in separate hard interactions, are responsible for a large portion of the structure in high-energy proton collisions. Together with initial and final state radiation, they are at the heart of generic Monte-Carlo generators, and necessary in order to obtain realistic descriptions of the collisions and reach agreement with data. There is ongoing effort to take the description of multiparton interactions from a model and tune base to a description resting on a solid foundation within QCD \cite{Diehl:2015bca,Manohar:2012jr,Diehl:2011tt,Blok:2010ge,Bansal:2014paa}.

Factorization for double parton scattering, the most frequent type of hard multiparton interactions, has been investigated in great detail \cite{Diehl:2015bca,Manohar:2012jr,Diehl:2011yj,Diehl:2011tt}, although there are still a few stones left to be turned for a proof at the same level of rigor as those for single Drell-Yan or Higgs production. 

The DPS cross section is given in terms of two hard, short distance interactions calculable in perturbation theory, and double parton distributions (DPDs) describing the partons in the incoming proton states. There is a large set of DPDs describing not only the different parton types and their kinematical distribution, but also correlations between for example color and spin of the two partons \cite{Manohar:2012jr,Diehl:2011yj,Mekhfi:1985dv}. These distributions however, are largely unknown. The flora of unknown non-perturbative distributions is currently the major hurdle for an accurate description of DPS cross sections. Although the distributions themselves cannot be calculated, their evolution can. Together with bounds on the distributions from positivity \cite{Diehl:2013mla,Kasemets:2014yna} this already provides strong limitations on when and where the different distributions can be of phenomenological importance and leave sizable footprints in the data. For example, the color correlations are suppressed for processes at large scales  \cite{Manohar:2012jr,Mekhfi:1988kj}, and different spin correlations, in particular for linearly polarized gluons at small momentum fractions, are also suppressed at high or moderate scales \cite{Diehl:2014vaa}. On the contrary, quark polarizations only slowly get washed out by evolution and can therefore remain sizable up to high scales.
 
 From a different perspective, a few model calculations have already been done for quark and anti-quark DPDs \cite{Rinaldi:2014ddl,Rinaldi:2013vpa,Chang:2012nw,Broniowski:2013xba}. In 
\cite{Golec-Biernat:2015aza}  two-gluon correlators have been modeled in the limit of zero relative transverse momentum using a parametrization of single gluon distributions. There are a few model studies in the literature so far for the spin correlations of two quarks in DPDs. When both quarks are longitudinally polarized, the correlations are calculated in a light-front constituent quark model in \cite{Rinaldi:2014ddl}. In \cite{Chang:2012nw}, two quark correlators  have been calculated for longitudinal and transverse polarization  and their relative magnitude is estimated in a bag model. In \cite{Broniowski:2016trx},  the unpolarized quark and gluon DPDs have been studied in a  valon model by factorizing  the  transverse momentum dependence and writing the rest using constraints from single parton distributions. However, to the best of our knowledge, there has been no model calculations dealing with the spin correlations in DPDs of the mixed quark-gluon state. The quark-gluon DPDs contribute to several interesting DPS processes, such as the production of a $W$ boson in association with a heavy meson \cite{Baranov:2016mix,Leontsinis:2016bsl}, a vector boson and dijet \cite{Aad:2013bjm,Chatrchyan:2013xxa} and even a Higgs boson in combination with a vector boson.

 One of the approaches to calculate the DPDs is to use the light-front wave function framework. The proton state is expanded in Fock space in terms of multi-parton light-front wave functions (LFWFs)\cite{Brodsky:1997de}. The valence sector gets contributions from the three quark wave function, but it is only the higher Fock components
 that have a gluon as one of the constituents. The LFWFs of the proton are boost invariant non-perturbative objects and need to be modeled.  Due to the complexity in modeling the higher Fock components, most of the model calculations in terms of the LFWFs are restricted to the valence quark sector. In this work, we study the quark-gluon DPDs  by replacing the proton target state by a simpler state, a quark dressed with a gluon. This can be thought of as a simple composite and relativistic spin $1/2$ state, having a gluon degree of freedom \cite{Harindranath:1998ve}.  The two-particle boost invariant LFWF can be calculated perturbatively, so in this sense it can be thought of as a perturbative model, which is based on field theory. Indeed, this model gives an intuitive picture of the distribution functions \cite{Mukherjee:2015aja, Mukherjee:2014nya}, as it has close connection with the parton model, but the partons, namely the quark and the gluon, have transverse momenta and interact. 
 
In the present study, we put emphasis on the different spin correlations between a quark and a gluon inside an unpolarized state as well as the differences between the kinematical distributions of the polarized and unpolarized partons. There are several studies on the effects which spin correlations between the two partons can have on DPS cross sections \cite{Echevarria:2015ufa,Kasemets:2012pr,Manohar:2012jr,Mekhfi:1983az}.  The polarized distributions can be divided into two types, longitudinal and transverse/linear polarization, where transverse is for quarks and linear is for gluons. While longitudinal polarization affects the total rate of DPS and the distributions in for example rapidity and transverse momentum, the transverse/linear polarization leads to dependences on the azimuthal angles between particles produced in the two different hard interactions, for example, the outgoing leptons in double Drell-Yan.

The outline of the paper is as follows: In section~\ref{sec:overlap} we discuss the DPDs for a quark and a gluon and their overlap representation in the light-front dressed quark model. We describe the calculation of the different DPDs in section~\ref{sec:calc}, and present numerical results in section~\ref{sec:num}. Conclusions are given in section~\ref{sec:concl}.

%%%%%%%%%%%%%%%%%%%%%%%%%%%%%%%%%%%%%%%%%%%%%%%%%%%%%%%%%%%%%%%%%%%%
\section{Overlap representation of the double parton distributions}
\label{sec:overlap}
%%%%%%%%%%%%%%%%%

%%%%%%%%%%%%%%%%%
\subsection{Double parton distributions}
\begin{figure}
\begin{center}
\includegraphics[width=0.4\textwidth]{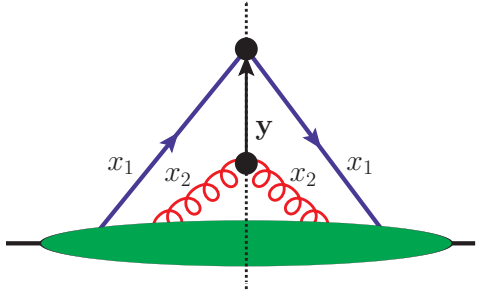}
\caption{\label{fig:dpd} Illustration of the DPD for a quark and a gluon.}
\end{center}
\end{figure}

%%%%%%%%%%%%%%%%%
The double parton distributions for quarks and gluons, illustrated in figure~\ref{fig:dpd}, are given by \cite{Diehl:2011yj}
\begin{align}
 F_{a_1a_2}(x_1,x_2,\vek{y})
 & = 2p^+ (x_1\ms p^+)^{-n_1}\, (x_2\ms p^+)^{-n_2}
        \int \frac{dz^-_1}{2\pi}\, \frac{dz^-_2}{2\pi}\, dy^-\;
           e^{i\ms ( x_1^{} z_1^- + x_2^{} z_2^-)\ms p^+}
\nonumber \\
 & \quad \times \left<p|\, \mathcal{O}_{a_2}(0,z_2)\,
            \mathcal{O}_{a_1}(y,z_1) \,|p\right> \,,
\end{align}
where $n_i = 1$ if parton number $i$ is a gluon, otherwise,  $n_i = 0$ and $\left|p\right>$ is a proton state with momentum $p$. $x_1$ and $x_2$ are momentum fractions of the partons and $\y$ is the relative
transverse distance between them. 

The quark operators are 
\begin{align}
\label{eq:quark-ops}
  \mathcal{O}_{a_i}(y,z_i)
   &= \bar{q}_i\bigl( y - \half z_i \bigr)\,
       \Gamma_{a_i} \, q_i\bigl( y + \half z_i \bigr)
   \Big|_{z_i^+ = y^+_{\phantom{i}} = 0,\; \vek{z}_i^{} = \vek{0}} \,,
\end{align}
where $\Gamma_{a_i}$ is the projection
\begin{align}
  \label{eq:quark-proj}
  \Gamma_q & = \half \gamma^+ \,,
&
  \Gamma_{\Delta q} &= \half \gamma^+\gamma_5 \,,
&
  \Gamma_{\delta q}^j = \half i \sigma^{j+} \gamma_5 \quad (j=1,2)
\end{align}
onto an unpolarized quark ($q$), longitudinally polarized quark ($\Delta
q$) or transversely polarized quark ($\delta q$). 

The operators for gluons are
\begin{align}
\label{eq:gluon-ops}
  \mathcal{O}_{a_i}(y,z_i)
   &= \Pi_{a_i}^{jj'} \, G^{+j'}\bigl( y - \half z_i \bigr)\,
        G^{+j}\bigl( y + \half z_i \bigr)
   \Big|_{z_i^+ = y^+_{\phantom{i}} = 0,\; \vek{z}_i^{} = \vek{0}}
\end{align}
with the projections
\begin{align}
  \label{eq:gluon-proj}
  \Pi_g^{jj'}  &= \delta^{jj'} \,,
&
  \Pi_{\Delta g}^{jj'} &= i\epsilon^{jj'} \,,
&
  [\Pi_{\delta g}^{kk'}]^{jj'} &= \tau^{jj'\!,kk'} \,,
\end{align}
onto unpolarized gluons ($g$), longitudinally polarized gluons ($\Delta
g$) and linearly polarized gluons ($\delta g$). We have not made explicit, the Wilson lines necessary to render the operators gauge invariant. For the color singlet DPDs these reduce to unity, but they are of importance for the color interference distributions. More details on the Wilson line structure in the DPDs can be found in \cite{Diehl:2011yj,Manohar:2012jr,Buffing:2016xx}.
The DPD correlator can be decomposed onto scalar DPDs describing different spin and spin-separation correlations. For a quark and a gluon the decomposition takes the form \cite{Diehl:2013mla}: 
\begin{align}
\label{eq:def-qg}
  F_{qg}(x_1,x_2,\y) & = f_{qg}(x_1,x_2,y) \,,
  \nonumber\\
  F_{\Delta q \Delta g}(x_1,x_2,\y) & =
     f_{\Delta q \Delta g}(x_1,x_2,y) \,,
  \nonumber\\
  F_{q \ms \delta g}^{jj'}(x_1,x_2,\y) & =
     \tau^{jj'\!,\y\y} M^2 f_{q \ms \delta g}(x_1,x_2,y) \,,
  \nonumber\\
  F_{\delta q \ms g}^j(x_1,x_2,\y) & =
     \ytilde^j M f_{\delta q \ms g}(x_1,x_2,y) \,,
  \nonumber\\
  F_{\delta q \delta g}^{j,kk'}(x_1,x_2,\y) & =
     \!\! {}- \tau^{\ytilde j, kk'} M f_{\delta q \delta g}(x_1,x_2,y)
  \nonumber\\
  & \quad - \bigl( \ytilde^j \tau^{kk'\!,\y\y}
                 + \y^j \tau^{kk'\!,\y\ytilde} \bigr) \,
              M^3 f_{\delta q \delta g}^t(x_1,x_2,y) \,.
\end{align}
Here $M$ is the mass of the proton, $\ytilde^j = \epsilon^{jj'} \y^{j'}$ and
$y=\sqrt{\y^2}$; $\tau^{jj'\!,\y\y} = \tau^{jj'\!,kk'}\, \y^k \y^{k'}$ and
\begin{align}
  \tau^{jj'\!,kk'} = \half \ms \bigl( \delta^{jk}\delta^{j'k'}
     + \delta^{jk'}\delta^{j'k} - \delta^{jj'}\delta^{kk'} \bigr) \,.
\end{align}

%%%%%%%%%%%%%%%%%
\subsection{Overlap representation}
%%%%%%%%%%%%%%%%%
As stated in the introduction, instead of a proton state, we take the incoming state to be a quark dressed with a gluon.
We will only consider distributions where the dressed quark is unpolarized, and thus average  over the helicities of the parent quark. The
state can be expanded in Fock space in terms of multi-parton light-front wave
functions (LFWFs). For our calculation, we keep only the two-parton wave
function. We write the state as \cite{Harindranath:1998ve} 
\be \label{dqs}
  \Big{| }p^{+},p_{\perp},s  \Big{\rangle} &=& \Phi^{s}(p)
b^{\dagger}_{s}(p)
 | 0 \rangle +
 \sum_{s_1 \lambda} \int \frac{dx_1 d^2 q_{1\perp}}{\sqrt{x_1}}\int \frac{dx_2 d^2 q_{2\perp}} {\sqrt{x_2}} 
 \delta(1-x_1-x_2)\\ \nn &&\quad \times { 1 \over \sqrt{16 \pi^3 }}
\delta^2(p_\perp - q_{1 \perp}-x_1 p_\perp -q_{2\perp} -x_2 p_\perp  ) \psi^s_{s_1,\lambda} (x_1,q_{1\perp}) 
b^{\dagger}_{s_1}(p_1)
 a^{\dagger}_{\lambda}(p_2)  | 0 \rangle\,.
\ee
$ \Phi^{s} $ is the  single particle (quark)  LFWF and  $\psi^s_{s_1,\lambda} (x_1,q_{1\perp})$ 
 the two particle (quark-gluon) boost invariant LFWF, 
 $s_1$ and $\lambda$ are the helicities of the quark and gluon in the
dressed quark respectively while $s$ is the helicity of the 
dressed quark.  $ \Phi^{s}(p)$ gives  the normalization of the state,
which upto the order we are calculating is unity. 

The square of the wave function $ \psi^{s}_{s_1 \lambda}$ gives the
probability to find a bare
quark and a bare gluon in the dressed quark. The Jacobi momenta $(x_i, q_{i \perp})$ are defined as  :
\be
p_i^+= x_i p^+, ~~~~~~~~~~p_{i \perp}= q_{i \perp}+x_i p_\perp \,,
\ee
where $\sum_i x_i=1$, $\sum_i q_{i\perp}=0$.

These two-particle LFWFs can be calculated perturbatively, and an analytic
expression is obtained  \cite{Harindranath:1998ve}:
\be \label{tpa}
\psi^{s a}_{s_1 \lambda}(x,q_{\perp}) =
\frac{1}{\Big[    m^2 - \frac{m^2 + (q_{\perp})^2 }{x} -
\frac{(q_{\perp})^2}{1-x} \Big]}
\frac{g}{\sqrt{2(2\pi)^3}} T^a \chi^{\dagger}_{s_1}
\frac{1}{\sqrt{1-x}}
\nn \\ \Big[
-2\frac{q_{\perp}}{1-x}   -
 \frac{(\sigma_{\perp}.q_{\perp})\sigma_{\perp}}{x}
+\frac{im\sigma_{\perp}(1-x)}{x}\Big]
\chi_s (\epsilon_{\perp \lambda})^{*}.
\ee
Here we have used the two component formalism \cite{Zhang:1993dd}; $\chi$ is the two
component spinor, $T^a$ are the color $SU(3)$ matrices, $m$ is
the mass of the quark and $\epsilon_{\perp \lambda}$ is the polarization
vector of the gluon; $\perp=1,2$.  We have taken the mass of the dressed
quark to be the same as the mass of the bare quark \cite{Harindranath:1998pd,Harindranath:1998pc}. 
This state can be thought of as a simplified model for the bound state of a spin-1/2 particle
and a spin-1 particle. 

Using the Fock space expansion above, $F(x_1,x_2,\y)$ can be written in terms
of an overlap of the two-particle LFWFs. 
Contribution to $F(x_1,x_2,\y)$ comes from two terms. The first term is of
the form 
\be\label{eq:unphy}
F_{s_1 s_1', \lambda \lambda'} (x_1,x_2,\y) &=& { 1\over  8 \pi^2 } \sum_{s} i^{(s_1-s_1')} \int  d^2 \tilde
{\vek{q}} \int d^2 \vek{q}  \delta(x_1-1-x_2) \nonumber \\&& ~~~\psi^{*,s}_{s'_1,\lambda'}(1+x_2,\tilde {\vek{q}})
\psi^s_{s_1, \lambda}(x_1,\vek{q}) 
\frac{1-x_1}{x_2} e^{-i
(\tilde {\vek{q}} - \vek{q})\cdot \vek{y}}. 
\ee
The delta function forces $x_1=1+x_2$, but as both $x_1 $ and $x_2$ are positive and less that
$1$, this term cannot contribute to the DPDs. Equation \eqref{eq:unphy} can be interpreted as the distribution of a quark and an anti-gluon or alternatively, as describing the situation where the incoming quark absorbs (rather than emits) a gluon. The contribution from the second term gives, 
\be
F_{s_1 s_1', \lambda \lambda'} (x_1,x_2,\y) &=& { 1\over  8 \pi^2 } \sum_{s} i^{(s_1-s_1')} \int  d^2 \tilde
{\vek{q}} \int d^2 \vek{q} \delta(1-x_1-x_2) \psi^{*,s}_{s'_1,\lambda'}(x_1,\tilde {\vek{q}})
\psi^s_{s_1, \lambda}(x_1,\vek{q}) \nonumber\\&&~~~~~~~~~~~~~~~
\frac{1-x_1}{x_2} e^{-i
(\tilde {\vek{q}} -\vek{q})\cdot \y}. 
\ee
This term has the correct support property and $x_2=1-x_1$, and thus contribute to the 
physical region of the DPDs in the dressed quark model. Note however that this requirement leads to heavily constrained kinematics for the two partons, as expected in this simple two-particle model. In order to study non-trivial correlations between the momentum fractions $x_1$ and $x_2$, higher Fock states would have to be included. However, as we will demonstrate below, already at this order we can investigate the transverse momentum and spin correlations between the two partons in the different DPDs contributing to DPS cross sections.

%%%%%%%%%%%%%%%%%
\section{Calculation of the DPDs}
\label{sec:calc}
%%%%%%%%%%%%%%%%%
Summing over the helicity states of the quark and gluon, with equal helicities in amplitude and conjugate, gives the unpolarized DPD
\be
f_{qg}(x_1,x_2,y) &=& {1\over 8\pi^2}\sum_{s,s_1,\lambda}  \int  d^2 \tilde
{\vek{q}} \int d^2 {\vek{q}} \delta(1-x_1-x_2)
 e^{-i(\tilde {\vek{q}} -{\vek{q}})\cdot \y} 
 \nn\\&&
 \quad \times\psi^{*,s}_{s_1,\lambda}(x_1,\tilde {\vek{q}})
\psi^s_{s_1, \lambda}(x_1,{\vek{q}})\,.
\ee
For the longitudinally polarized DPD we get
\be
f_{\Delta q \Delta g}(x_1,x_2,y) &=& {1\over 8\pi^2}\sum_{s,s_1,\lambda}
2 s_1 \lambda   \int  d^2 \tilde
{\vek{q}} \int d^2 {\vek{q}} \delta(1-x_1-x_2)
 e^{-i(\tilde {\vek{q}} -{\vek{q}})\cdot \y} 
 \nn\\&&
 \quad \times\psi^{*,s}_{s_1,\lambda}(x_1,\tilde {\vek{q}})
\psi^s_{s_1, \lambda}(x_1,{\vek{q}})\,,
\ee
where $(2s_1\lambda)$ equals $+1$ when the helicities of the two partons are aligned and $-1$ otherwise.
Using the light-front wave functions given above, these can be expressed in
terms of a few master integrals:
\begin{align}
I_1=\int d^2 \vek{k}  e^{i \y \cdot  \vek{k} }\frac{1}{{\vek{k}}^2+\Lambda^2} & 
= 2\pi K_0 (\Lambda y) \,,\nn\\
I_2=\int d^2 \vek{k}  e^{i \y \cdot  \vek{k} }\frac{|\vek{k} |\cos\phi }
{{\vek{k}}^2+\Lambda^2} & = 2\pi i \cos\phi_y \Lambda K_1(\Lambda y) \,,\nn\\
I_3=\int d^2 \vek{k}  e^{i \y \cdot  \vek{k} }\frac{|\vek{k}
|\sin\phi}{{\vek{k}}^2+\Lambda^2} & = 2\pi i \sin\phi_y \Lambda K_1(\Lambda y) \,,
\end{align}
where $K_n$ is the n-th hyperbolic Bessel function of the second kind. 
$\Lambda=m (1-x_1)$, $\phi$ is the azimuthal angle of $\bf{k}_\perp$ and $\phi_y$ the azimuthal angle of $\bf{y}$. 
We get the unpolarized DPD :
\be
f_{qg} (x_1,x_2,y) &=& {C_f \over 2 \pi^2} {g^2 \over 2 {(2 \pi)}^3} 
 \delta (1-x_1-x_2) {1 \over (1-x_1)} 
\nn\\
&&\quad\times \Big [ (1-x_1)^4 m^2 I_1^2 -
(1+x_1^2) (I_2^2 + I_3^2) \Big ]
\nonumber\\
& = &{\alpha_s  C_f \over 2\pi^2}  \delta (1-x_1-x_2) \Big [ (1-x_1)^3 m^2 K_0^2 ((1-x_1) m y) 
\nn\\
&&\quad\quad + (1+x_1^2)
(1-x_1) m^2 K_1^2 ((1-x_1) m y)\Big ] \, .
\ee 
The DPD for longitudinally polarized quark and gluon is given by :
\be
f_{\Delta q\Delta g} (x_1,x_2,y) &=& {1\over 2\pi^2} C_f {g^2 \over 2 {(2 \pi)}^3} \delta (1-x_1-x_2) {1 \over (1-x_1)} 
\nn\\
&& \quad \times \Big [ -(1-x_1)^4 m^2 I_1^2 -
(1-x_1^2) (I_2^2 + I_3^2) \Big ]
\nonumber\\
&=& {\alpha_s \over 2\pi^2} C_f  \delta (1-x_1-x_2) \Big [- (1-x_1)^3 m^2 K_0^2 ((1-x_1) m y) 
\nn\\
&&\quad\quad  + (1-x_1^2)
(1-x_1) m^2 K_1^2 ((1-x_1) m y)\Big ] \,.
\ee 
We can notice that the change in the distribution in going from the unpolarized to longitudinally polarized case is only the additional minus signs of the first term in the square brackets and of the $x^2$ factor in the second. This reflects that while the unpolarized is a sum over the four helicity states of the quark and gluon, the longitudinally polarized distribution is given by the difference between the configuration where the helicities are aligned and where the helicities are anti-aligned. 
The linearly polarized gluon and transversely polarized quark are interferences between helicity states,
 i.e. the quark (gluon) in the amplitude no longer has the same helicity as the quark  (gluon) in the conjugate amplitude. The relation between the linearly/transversely polarized distributions to specific helicity states was worked out in \cite{Diehl:2013mla} 
%%%
\begin{align}
s'_1 \lambda' s_1 \lambda & = \{+++-\} : & \frac{-1}{4} e^{2i\phi_y} y^2 M^2 f_{q\delta g} \,,\nn\\
s'_1 \lambda' s_1 \lambda & = \{++-+\} : & \frac{-i}{4} e^{i\phi_y} y M f_{\delta q g} \,,\nn\\
s'_1 \lambda' s_1 \lambda & = \{++--\} : & \frac{-2i}{4} e^{3i\phi_y} y^3 M^3 f^t_{\delta q\delta g} \,,\nn\\
s'_1 \lambda' s_1 \lambda & = \{-++-\} : & \frac{-2i}{4} e^{i\phi_y} y M f_{\delta q\delta g} \,.\nn\\
\end{align}
%%%
We therefore only need to calculate a single helicity state (sum only over the dressed quark helicity)
 for each different polarized distribution. We readily obtain for the distribution of an unpolarized quark and a linearly polarized gluon
\be
f_{q\delta g} (x_1,x_2,y) &=& {1\over 2\pi^2} C_f {g^2 \over 2 {(2 \pi)}^3} 
 \delta (1-x_1-x_2) {1 \over (1-x_1)} \Big [ {(-4) e^{-2 i \phi_y} \over
y^2 M^2} \Big ]{x_1 \over 2} \nonumber\\ && \Big [  (-I_2^2+I_3^2)  -
 2i I_2 I_3  \Big ]\nonumber\\ &&
= {\alpha_s \over   2\pi^2} C_f  \delta (1-x_1-x_2) \Big [ {2 x_1 (1-x_1)\over y^2}\Big ]  K_1^2 (\Lambda y) \,.
\label{fqdeltag}
\ee 
The distribution of a transversely polarized quark in combination with an unpolarized gluon takes the form
\be
f_{\delta q g} (x_1,x_2,y) &=& {1\over 2\pi^2} C_f {g^2 \over 2 {(2 \pi)}^3}
\delta (1-x_1-x_2) (1-x_1)  \Big [ {(-2) e^{- i \phi_y} \over
y M} \Big ] (-i m) \nonumber\\ && \Big [  -I_1 (I_2 + i  I_3) \Big ]\nonumber\\ &&
= {\alpha_s \over  2 \pi^2} C_f  \delta (1-x_1-x_2) m \Big [ {2(1-x_1)^2 \over y}\Big ] K_0 (\Lambda y)  K_1 (\Lambda y) \,.
\ee 
Finally, the two distributions with both transverse and linear polarization are given by
\be
f_{\delta q \delta g} (x_1,x_2,y) &=& {1\over 2 \pi^2} C_f {g^2 \over 2 {(2 \pi)}^3} 
 \delta (1-x_1-x_2) x_1 (1-x_1)  \Big [ { e^{- i \phi_y} \over
y M} \Big ] (-i m) \nonumber\\ && \Big [  2I_1 I_2 +2i  I_1 I_2 \Big ]\nonumber\\ &&
= {\alpha_s \over  2 \pi^2} C_f  \delta (1-x_1-x_2) m \Big [ {2x_1 (1-x_1)^2 \over y}\Big ] K_0 (\Lambda y)  K_1 (\Lambda y)
\ee 
and
%%%
\begin{align}
f^t_{\delta q \delta g} = 0 \,.
\end{align}
 %%%
 That $f^t_{\delta q \delta g}$ equals zero is not surprising since the distribution describes a difference in helicity between amplitude and conjugate amplitude by 3 units \cite{Diehl:2013mla}, which cannot be generated by the single $1\rightarrow2$ splitting. For massless quarks the perturbative $1\rightarrow2$ splitting kernels are given in \cite{Diehl:2011yj,Diehl:2014vaa}. Expanding our result in the limit of $m\rightarrow 0$ agrees with these results, and leads for example to zero values for $m y f_{\delta q \delta g}$ and $m y f_{\delta q g}$.
 
 \subsection{Color correlation distributions}
 In addition to the different spin correlations discussed in the last section, the color of the two partons can be correlated. For a DPD of a quark and a gluon the correlator can be decomposed into a color singlet distribution and a symmetric and anti-symmetric octet  \cite{Diehl:2011yj}
\begin{align}\label{eq:col-mixed}
F_{jj'}^{aa'} = \frac{1}{N_c(N_c^2-1)} \Bigg[  \,\clr{1}F \delta^{aa'}\delta_{jj'} - \,\clr{A}F \sqrt{2}i f^{aa'c}t^{c}_{jj'} + \sqrt{\frac{2N_c^2}{N_c^2-4}} \; \clr{S}F d^{aa'c}t^c_{jj'}  \Bigg].
\end{align}
The coefficients are choosen such that all three distributions enter the production of color singlets, such as Drell-Yan and Higgs production, with equal weight and thus the size of the three color distributions directly reflect their phenomenological importance. In the present model, the color structure separates and results in simple overall factors. Taking the ratio of the different distribution, thus gives an overall color factor indicating the relative size of the color interference distributions
\begin{align}
\frac{\clr{A}F}{\clr{1}F} & = -\frac{3}{\sqrt{2}} \,, &
\frac{\clr{S}F}{\clr{1}F} & = \sqrt{\frac{5}{2}} \,.
\end{align}
These relations, saturates the most stringent color bound on double parton distributions in \cite{Kasemets:2014yna} for zero values of the parton type interference distributions.
 
 %%%%%%%%%%%%%%%%%
 \section{Numerical results}
 \label{sec:num}
 %%%%%%%%%%%%%%%%%
 We will now move to numerical examinations of the different correlations between the quark and gluon inside the dressed quark state.
 
 In order to simplify the notation we define rescaled distributions for a quark and a gluon as
 \begin{align}
  \label{dist-list}
  h_{ab} & =(yM)^c f_{ab}\,, 
  \end{align}
   where $c$ equals 0 for $f_{q g}$ and $f_{\Delta q \Delta g}$, 1 for $f_{\delta q g}$ and $f_{\delta q \delta g}$, 2 for $f_{q \delta g}$, and 3 for $f_{\delta q \delta g}^t$. The pre-factors compensate for the factors of $y$ and $M$ in the decomposition of the DPDs. 
 We define the ratios of polarized to unpolarized distributions
 \begin{align}
 R_{ab} = \frac{h_{ab}(x_1,x_2,y)}{f_{qg}(x_1,x_2,y)} \,.
 \end{align}
 Such ratios can, for example, be used in order to model the polarized DPDs in terms of the unpolarized in similar fashion to what was done with massless splitting kernels in \cite{Diehl:2014vaa,Echevarria:2015ufa}. We set $m=M=0.33$~GeV for all numerical evaluations. Notice that the ratios only depend on the product of $m$ and $y$.  We are not investigating the DPDs in a full phenomenological model of the proton but instead  trying to gain information of the momentum and spin correlations between the quark and a gluon in a two-particle state. Therefore, instead of the magnitude of each DPD, it makes more sense to plot the ratios in order to gain insight into the relative size of the spin correlations compared to the case where the quark and gluon are unpolarized.
 
\begin{figure}
\begin{center}
\subfigure[]{\includegraphics[width=0.4\textwidth]{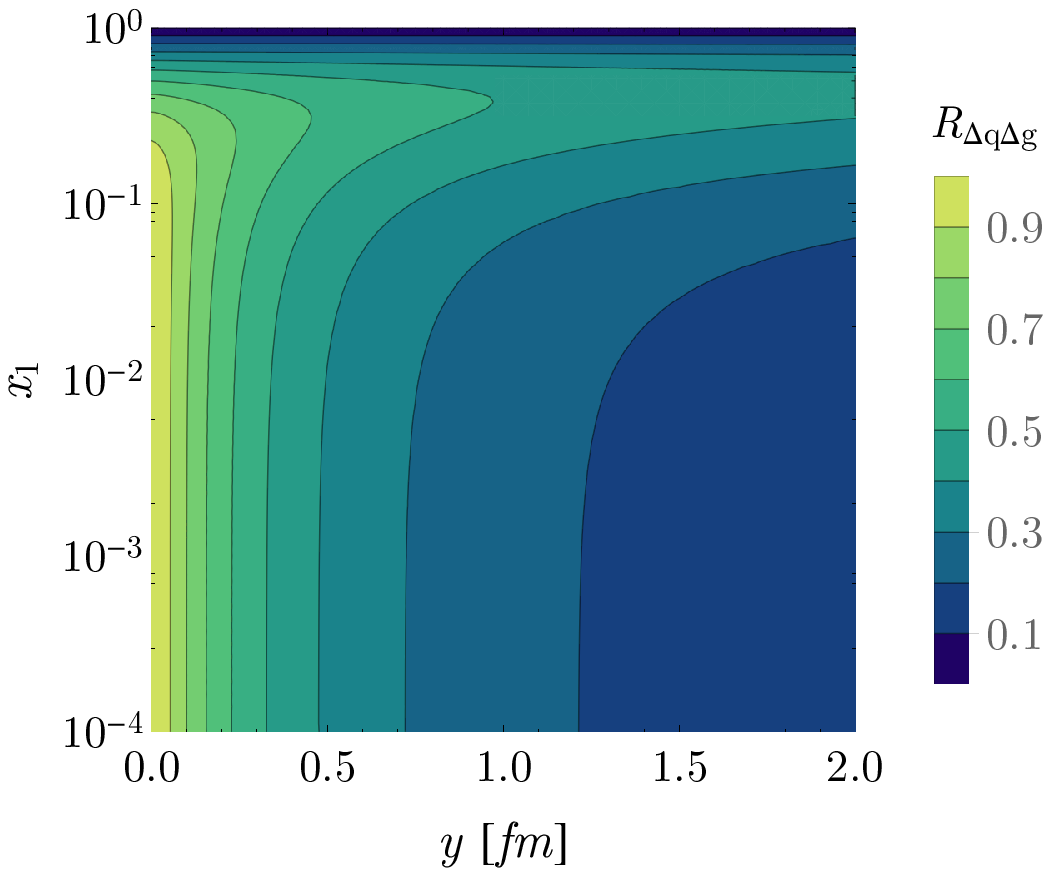}}
\hspace{3em}
\subfigure[]{\includegraphics[width=0.4\textwidth]{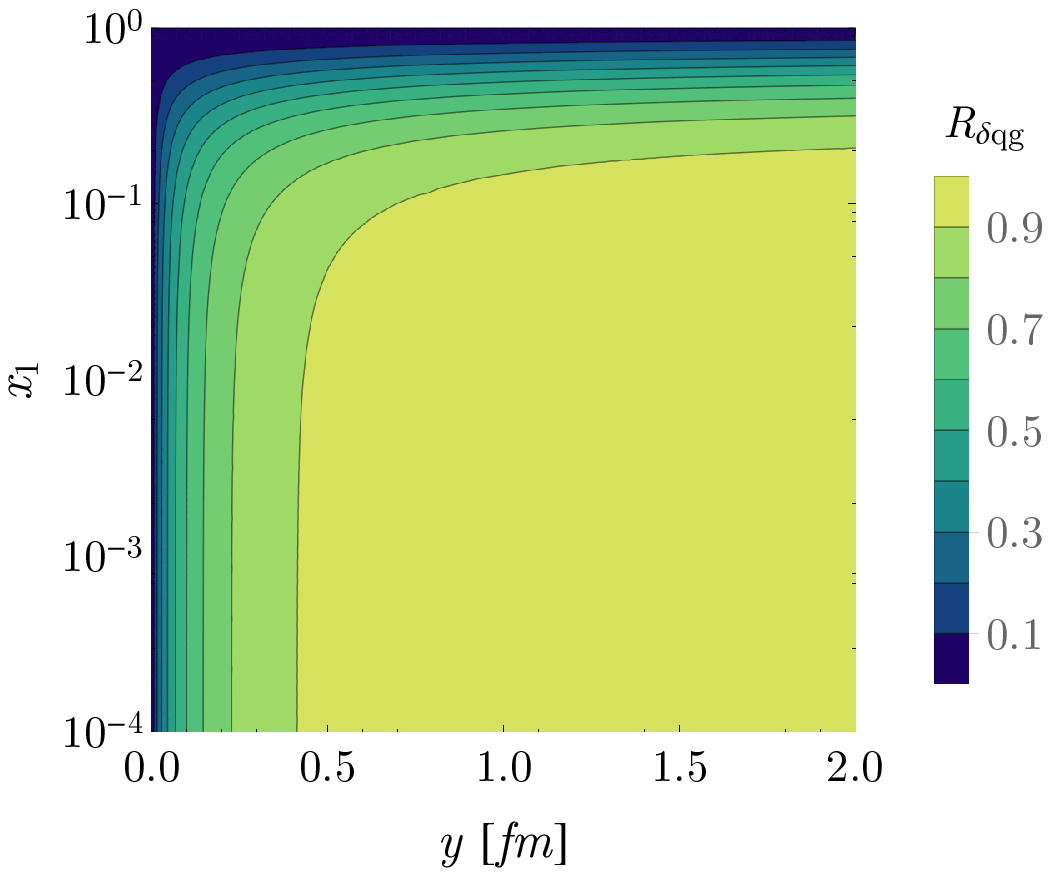}}
\\
\subfigure[]{\includegraphics[width=0.4\textwidth]{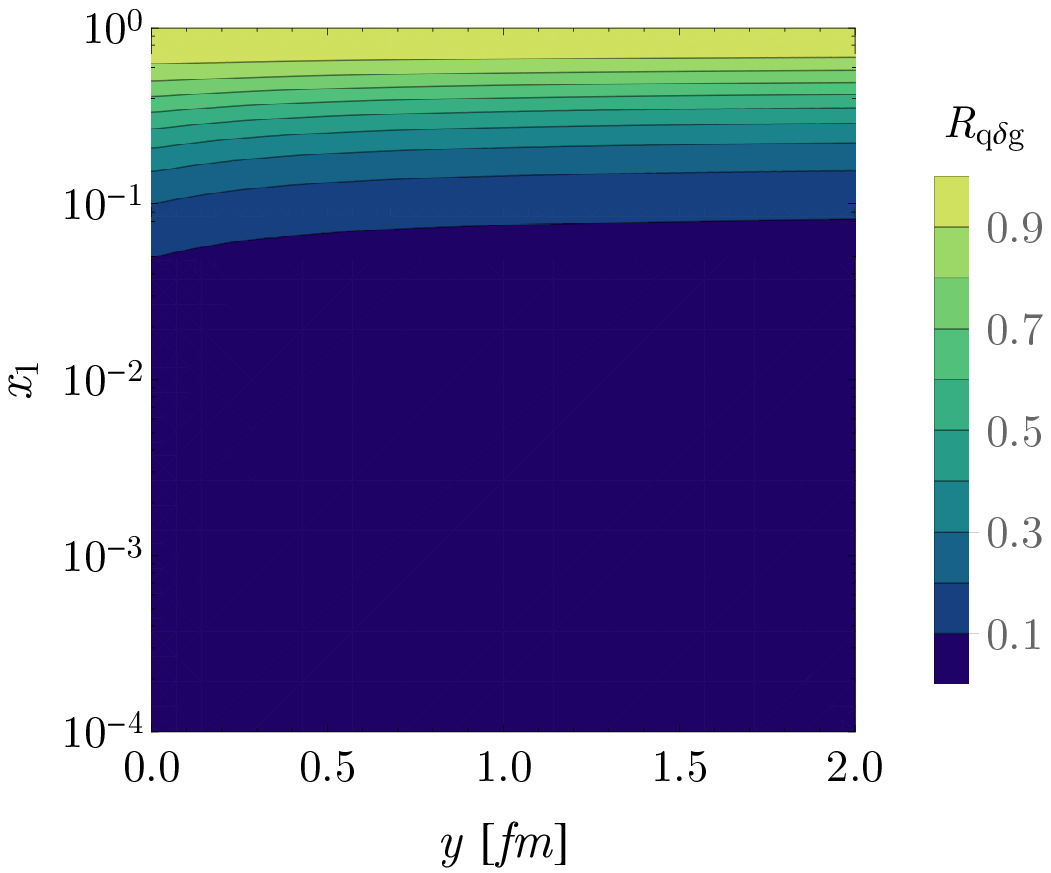}}
\hspace{3em}
\subfigure[]{\includegraphics[width=0.4\textwidth]{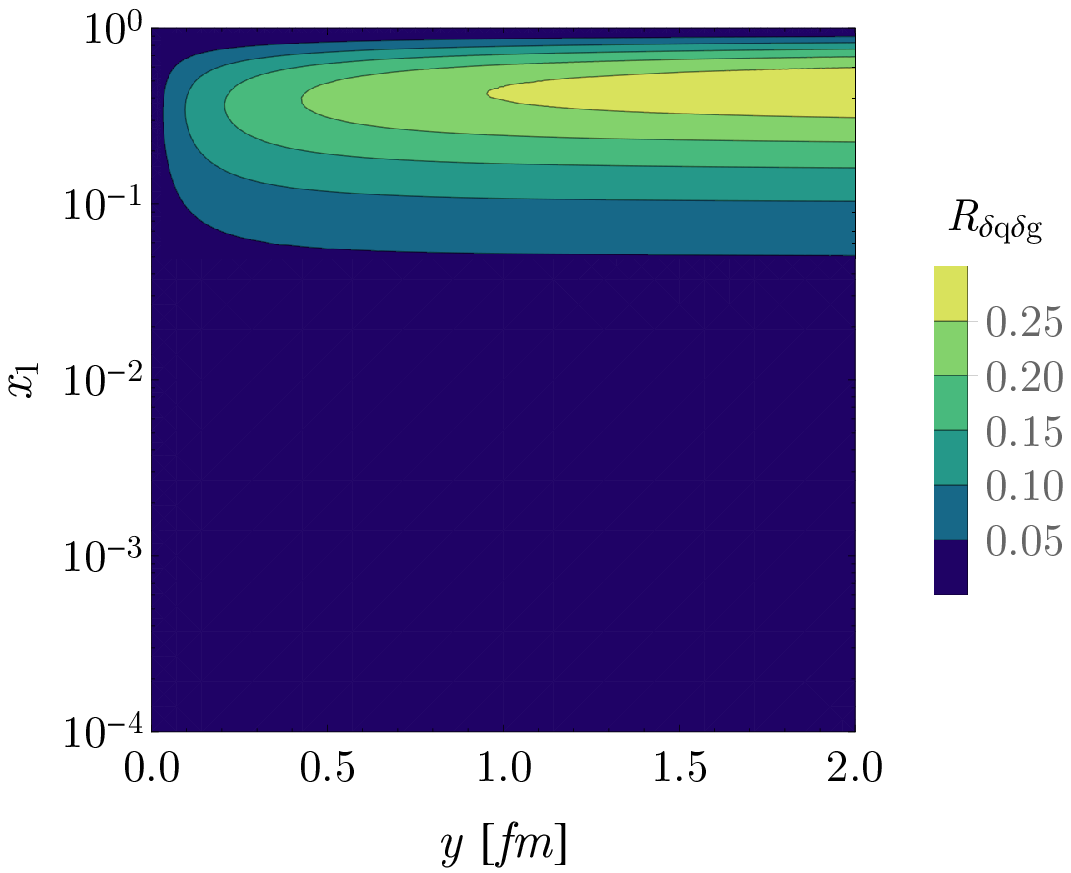}}
\caption{\label{fig:cont} $R_{\Delta q \Delta g}$, $R_{ q \delta g}$, $R_{\delta q g}$, $R_{\delta q \delta g}$ as a function of $x_1$ and $y$ with $m=0.33$~GeV. Notice the difference in color scale in (d).}
\end{center}
\end{figure}

\begin{figure}
\begin{center}
\subfigure[]{\includegraphics[width=0.4\textwidth]{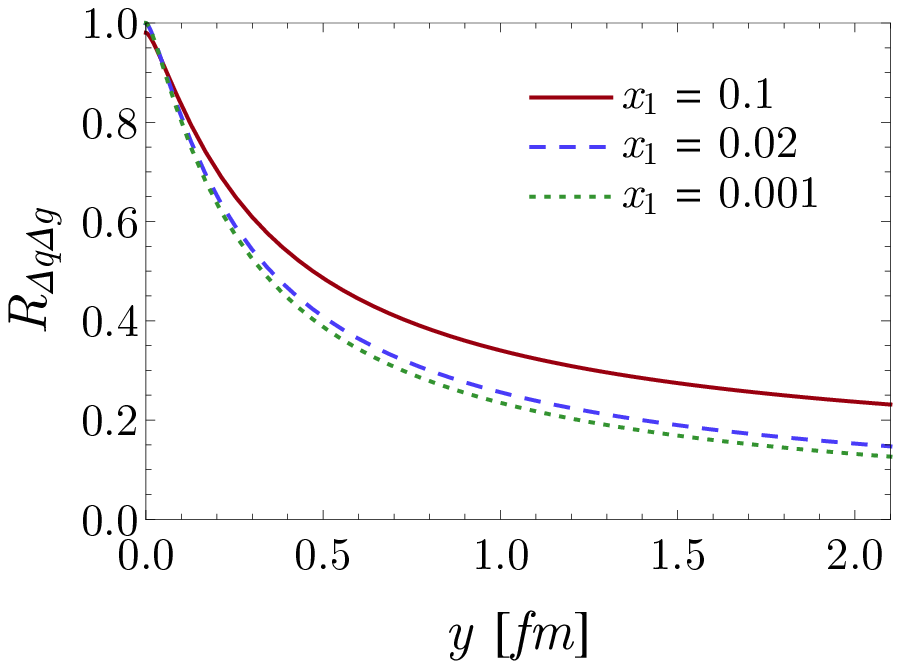}}
\hspace{3em}
\subfigure[]{\includegraphics[width=0.4\textwidth]{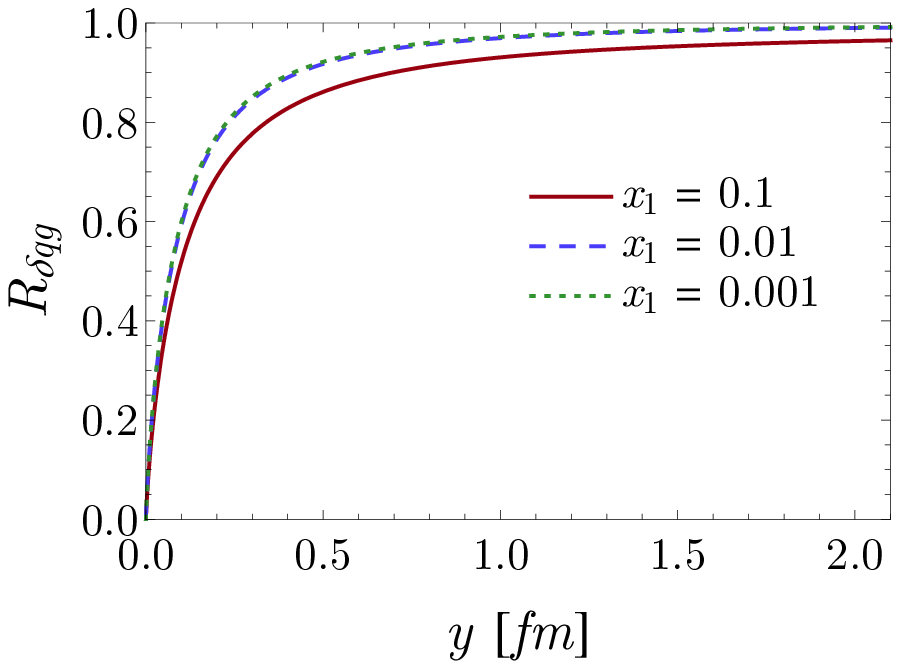}}
\\
\subfigure[]{\includegraphics[width=0.4\textwidth]{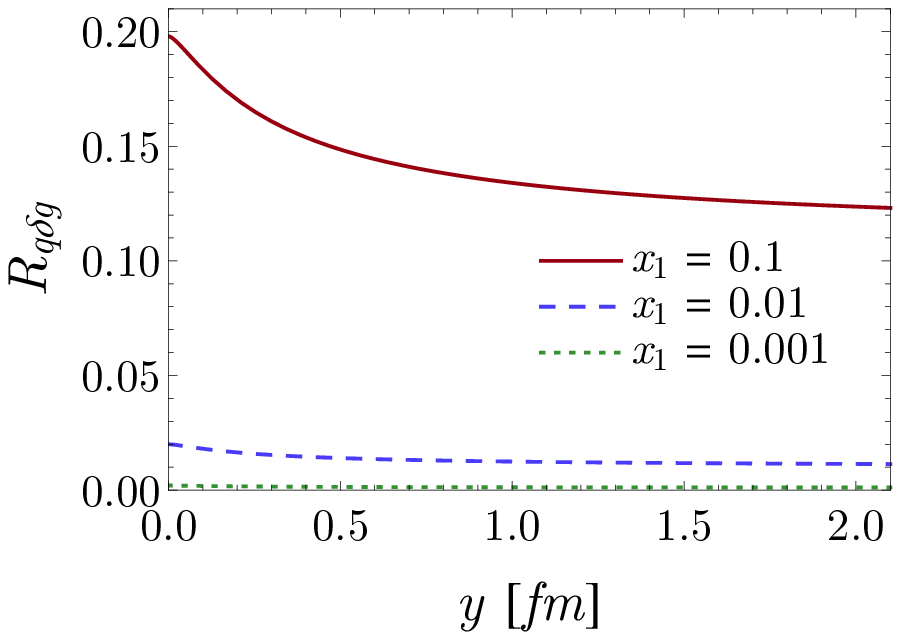}}
\hspace{3em}
\subfigure[]{\includegraphics[width=0.4\textwidth]{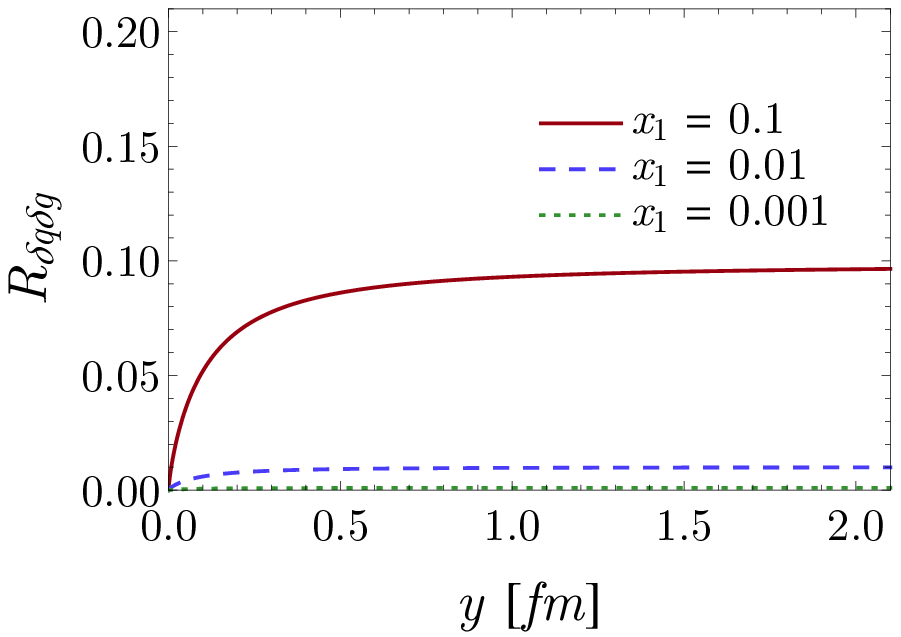}}
\caption{\label{fig:ratio_y} $R_{\Delta q \Delta g}$, $R_{\delta q g}$, $R_{ q \delta g}$, $R_{\delta q \delta g}$ as a function of $y$ for different $x_1$ fractions and $m=0.33$~GeV.  Notice the different $x_1$ fractions in (a).}
\end{center}
\end{figure}

\begin{figure}
\begin{center}
\subfigure[]{\includegraphics[width=0.4\textwidth]{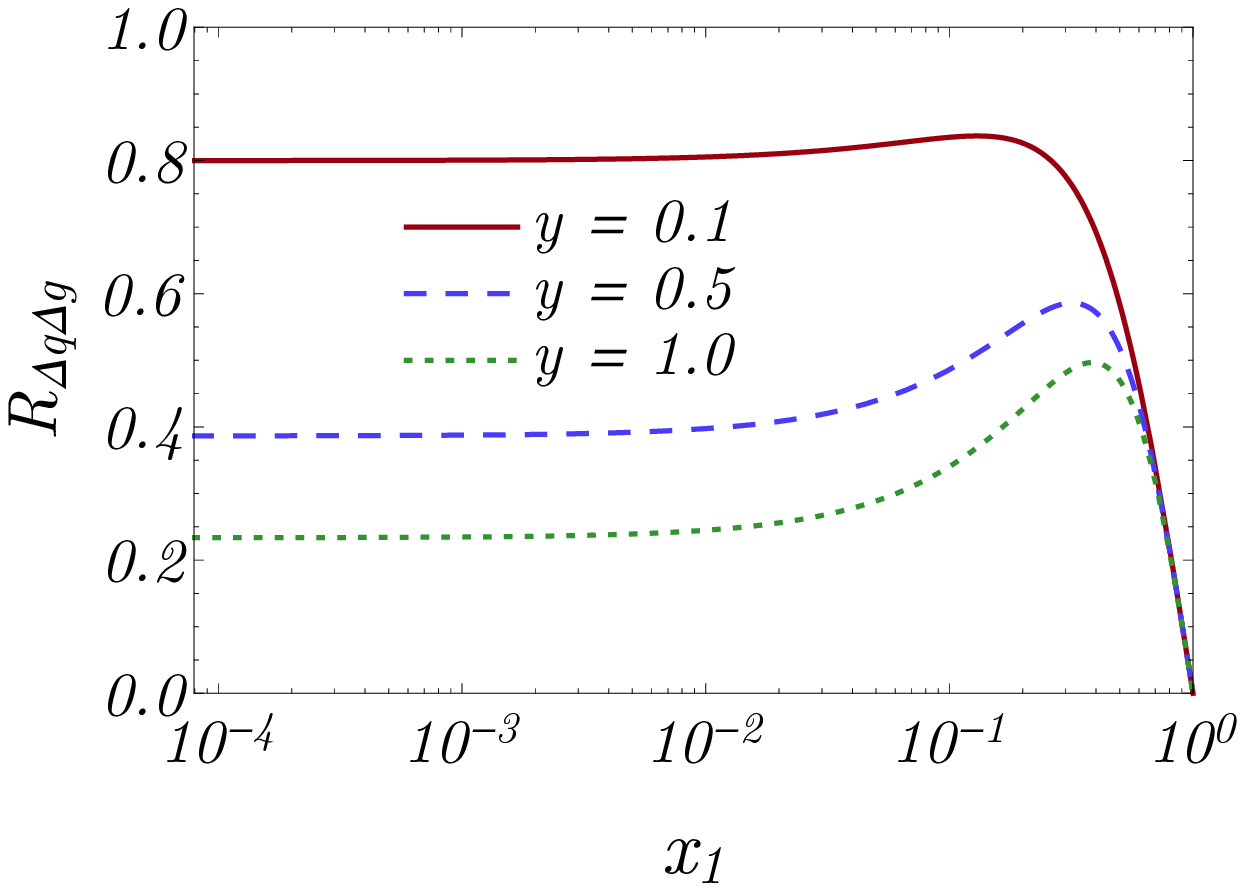}}
\hspace{3em}
\subfigure[]{\includegraphics[width=0.4\textwidth]{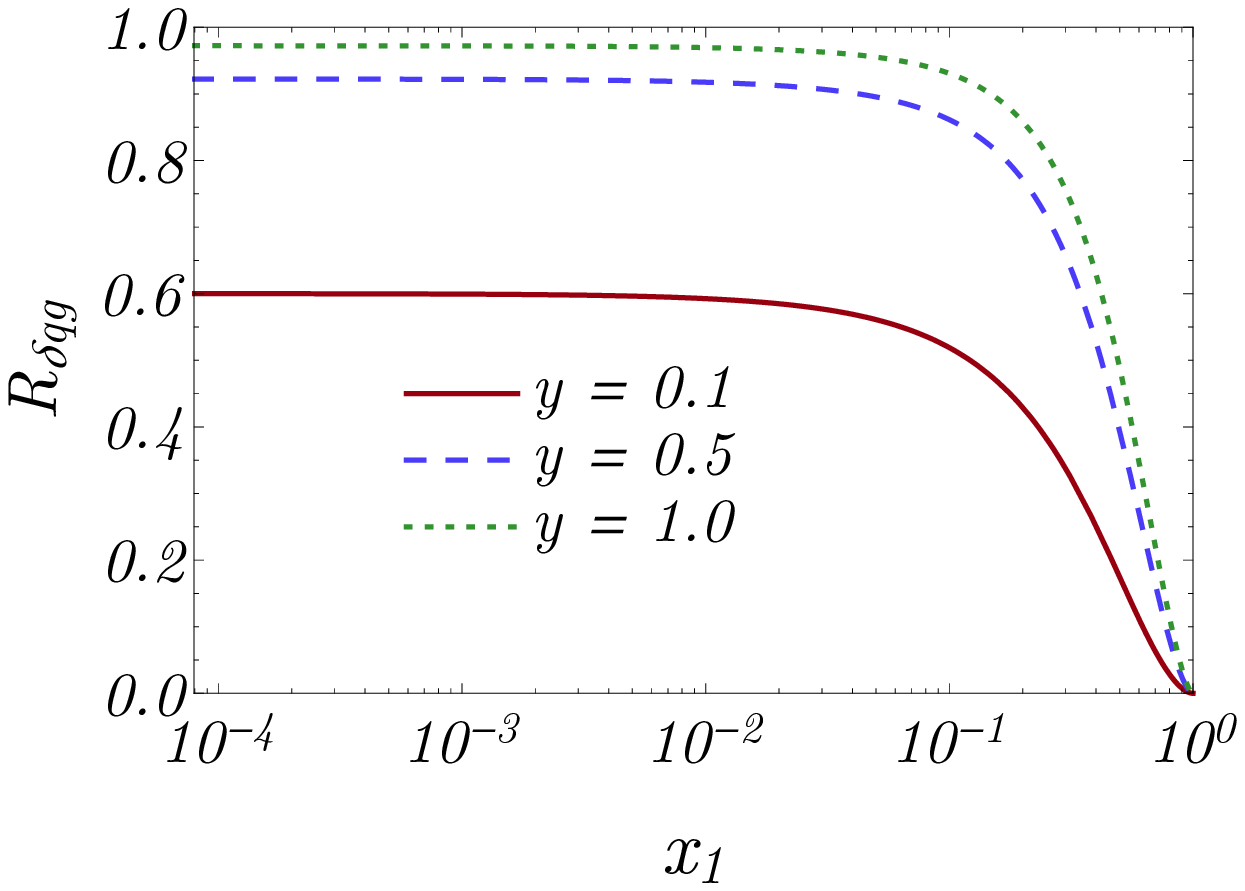}}
\\
\subfigure[]{\includegraphics[width=0.4\textwidth]{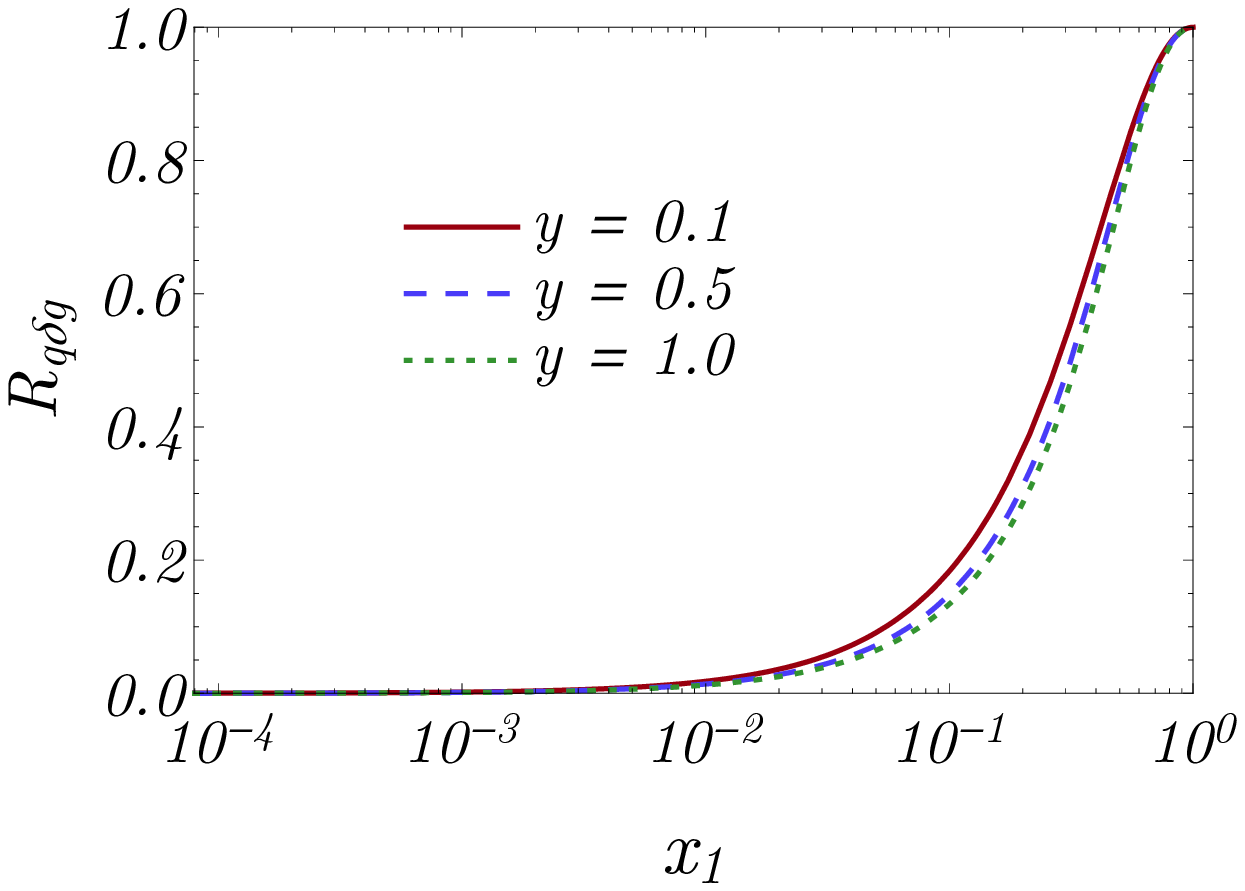}}
\hspace{3em}
\subfigure[]{\includegraphics[width=0.4\textwidth]{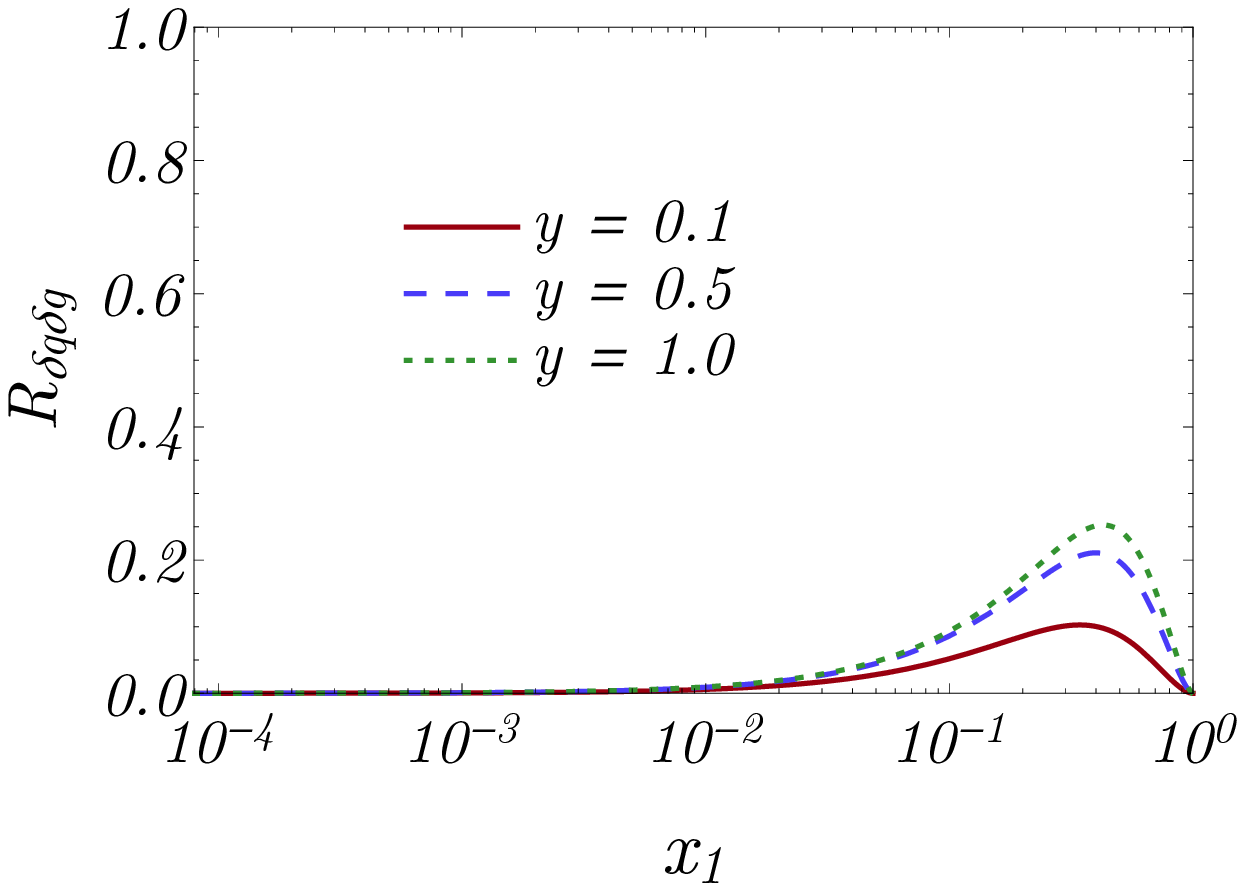}}
\caption{\label{fig:ratio_x}$R_{\Delta q \Delta g}$, $R_{\delta q g}$, $R_{ q \delta g}$, $R_{\delta q \delta g}$ plot 
against $x_1$ for different values of $y$ in fm and $m=0.33$~GeV.  }
\end{center}
\end{figure}

 Figure~\ref{fig:cont} shows the ratios for the different non-zero contributions in a contour plot as a function of the momentum fraction $x_1$ and the transverse distance $y$. The longitudinally polarized DPD is relatively large in a sizable region of $(x,y)$-space. The ratio is less than one, as expected from positivity bound, also observed for the double quark correlations in the constituent quark model \cite{Rinaldi:2014ddl}. The ratio $R_{\Delta q \Delta g}$  increases towards small $y$ and demonstrates non-trivial $x_1$-$y$ correlations, where the ratio is significantly broader (in transverse space) at larger $x_1$. The ratio with a transversely polarized quark, $R_{\delta q g}$ is large over a large area, but in contrast to the longitudinally polarized one $R_{\delta q g}$ goes to zero for small $y$. The $x_1$ dependence away from the small $y$ and large $x_1$ region is rather mild. Both ratios with a linearly polarized gluon peaks at larger $x_1$ and is small at small momentum fractions. Linear gluon polarization in combination with an unpolarized quark, $R_{q \delta g}$, can remain sizable down to $x_1\sim 0.1$ while the combination of a linear gluon and a transverse quark, $R_{\delta q \delta g}$, leads to a small ratio over the entire region. 
We further show the ratios as a function of $y$ for a few values of $x_1$ in figure~\ref{fig:ratio_y} and as a function of $x_1$ for a few values of $y$ in figure~\ref{fig:ratio_x}. These ratios clearly demonstrate the relative sizes of the polarized distributions and the $x_1$-$y$ correlations in the ratios. Figure~\ref{fig:ratio_y} shows the reduction in width of $R_{\Delta q \Delta g}$ towards smaller momentum fractions. The specific $x_1$ value does not much influence the $y$ dependence of $R_{\delta q g}$ while the linearly polarized gluons are very close to zero for the smaller momentum fractions. From figure~\ref{fig:ratio_x} we can see the nontrivial both $x_1$ shape and size dependence of $R_{\Delta q \Delta g}$ on the $y$ value, while, $R_{\delta q g}$ and $R_{\delta q \delta g}$ show that the different $y$ values lead to different sizes but similar $x_1$ dependence. $R_{g \delta q}$ remains constant for the $y$ values probed. $R_{\delta q \delta g}$  is small in the entire region, and the shape resembles that of the product of $R_{q\delta g}$ and $R_{\delta q g}$. From the plots one can see that  $R_{q\delta g}$ approaches $1$ as $x_1 \rightarrow 1$. This is expected from the analytic results, namely Eq. (\ref{fqdeltag}). However, for gluon dependent correlators in particular and DPDs in general, it is the small $x$ region that is more relevant phenomenologically.

\begin{figure}
\begin{center}
\subfigure[]{\includegraphics[width=0.4\textwidth]{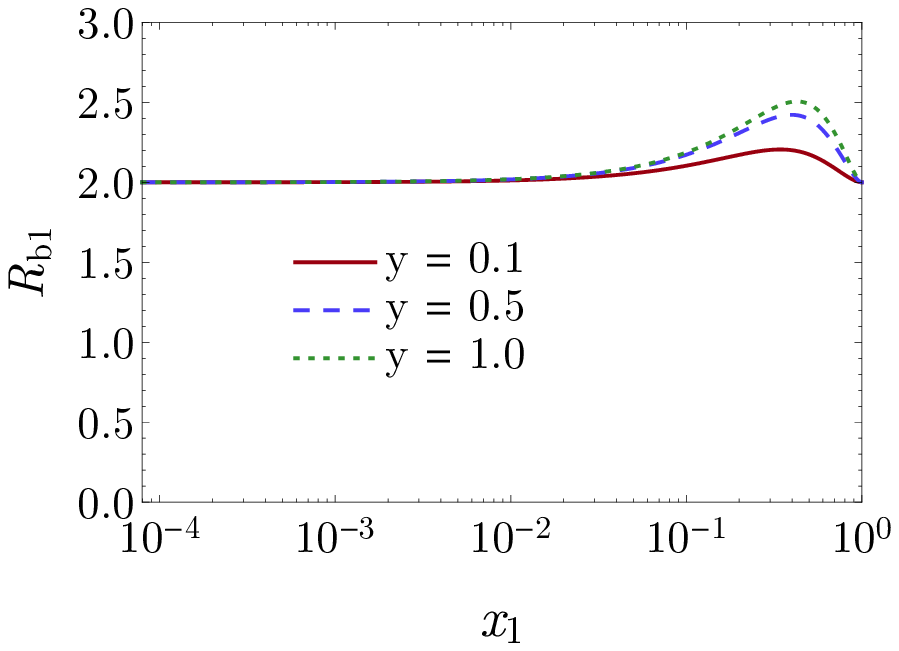}}
\hspace{3em}
\subfigure[]{\includegraphics[width=0.4\textwidth]{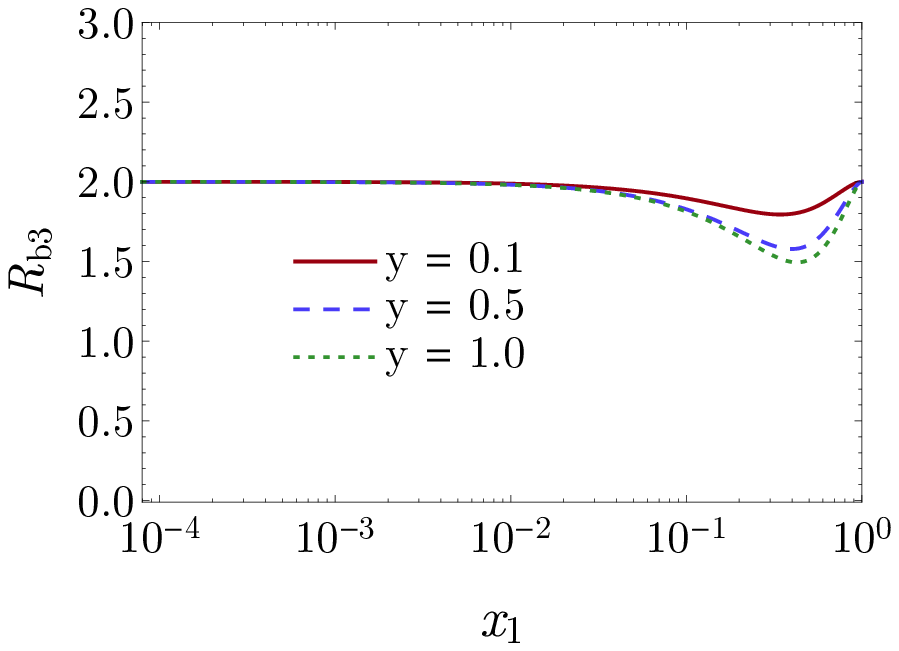}}
\caption{\label{fig:bounds} Eigenvalue of the helicity matrix normalized by the unpolarized distributions a function of against $x_1$ with $m=0.33$~GeV. The other two eigenvalue bounds are saturated.}
\end{center}
\end{figure}

We will now turn to investigate the positivity bounds derived in \cite{Diehl:2013mla}, from the positivity of the eigenvalues of the helicity density matrix,
 \begin{align}
   f_{ab} + h_{\delta a \delta b} - h_{\delta a \delta b}^t   
      & \pm \sqrt{(h_{\delta a \ms b} +  h_{a \ms \delta b})^2 +
        (f_{\Delta a \Delta b} - h_{\delta a \delta b}
       - h_{\delta a \delta b}^t)^2}  \,\geq\, 0 \,,
  \nonumber\\
  f_{ab} - h_{\delta a \delta b} + h_{\delta a \delta b}^t 
      & \pm \sqrt{(h_{\delta a \ms b} -  h_{a \ms \delta b})^2 +
        (f_{\Delta a \Delta b} + h_{\delta a \delta b}
       + h_{\delta a \delta b}^t)^2}  \,\geq\, 0 \,.
\end{align}
Figure~\ref{fig:bounds} shows the $x_1$ dependence of the two bounds with positive signs for the square-roots (divided by the unpolarized distributions) $R_{b1}$ and $R_{b2}$. The model satisfies the bounds, and both $R_{b1}$ and $R_{b2}$ are flat in a large range of $x_1$ and $y$ and approximately equal to $2$. Interestingly, the combination of the polarized distributions in the most stringent bounds (i.e. when the square-roots enter with negative signs) approximately equals the unpolarized distribution and thus the model saturates the bounds.
%
 %%%%%%%%%%%%%%%%%
 \section{Conclusions}
 \label{sec:concl}
 %%%%%%%%%%%%%%%%%
 We have presented what is, to the best of our knowledge, the first model study of the double parton distributions for a quark and a gluon, where either or both of the partons can be polarized. Instead of a proton, we used  a simple two-body spin $1/2$ state of a quark dressed with a gluon. 
 The light-front dressed quark model results for the double parton distributions of a quark and a gluon demonstrates several sizable correlation effects. In particular, the longitudinally polarized distributions and the distribution of a transversely polarized quark and unpolarized gluon is large for large ranges in both the momentum fraction $x_1$ and transverse distance $y$. In addition, we have found that there are strong correlations between the momentum fraction $x_1$, the transverse distance and the relative (compared to the unpolarized) sizes of the spin correlations.  The model saturates the most stringent positivity bounds for both spin and color correlation distributions. The strong correlation between the size of polarized distributions and the transverse distance can be contrasted with the weak correlations with the transverse dependence found in previous model calculations \cite{Rinaldi:2014ddl,Rinaldi:2013vpa,Chang:2012nw}. However, as pointed out in \cite{Rinaldi:2014ddl}, the strength of these correlations is a model dependent feature related to the orbital angular momentum of the quarks.

The double parton distributions are objects of intense study in recent days. As enough experimental data is yet not available for a realistic estimate of these two-parton correlations, one often uses a factorized approximation for numerical estimates where one writes the DPDs in terms of products of single parton distributions or generalized parton distributions. Our simple model calculation for a dressed quark target does not support such factorized expressions and demonstrates several non-trivial and entangled correlations of the momentum and spin of the two partons. For example, the transverse distance dependence of the DPDs does not factorize. In order to examine more non-trivial $x_1$ and $x_2$ dependence, higher Fock state components must be included, which would also incorporate the scale dependence of the distributions.
 
 Our findings help pin down the structure and hierarchy of the many unknown non-perturbative double parton distributions and may aid in the creation of more realistic DPDs for input to phenomenological studies of DPS cross sections.
 
 %%%%%%%%%%%%%
\begin{acknowledgments}
We thank Piet Mulders for helpful discussions and comments, and also Jonathan Gaunt for his useful comments on the manuscript. This work is supported by the European Research Council under the "Ideas" program QWORK (contract 320389).
\end{acknowledgments}
%%%%%%%%%%%%%
 
\bibliographystyle{JHEP}
\bibliography{RefBib}
 
 \end{document}